# A Review of Light-Field Imaging in Biomedical Sciences

RUIXUAN ZHAO,[1][†] XUANWEN HUA,[2][†] WOONGJAE BAEK,[1] ZHAOQIANG WANG,[1] SHU JIA,[2] AND LIANG GAO[1],*

[1]*Department of Bioengineering, University of California Los Angeles, Los Angeles, CA 90095, USA*
[2]*Wallace H. Coulter Department of Biomedical Engineering, Georgia Institute of Technology and Emory University, Atlanta, Georgia 30332, USA*
[†]*These authors contributed equally to this paper*
*\*gaol@ucla.edu*

**Abstract:** Light-field imaging is an emerging paradigm in biomedical optics, offering the unique ability to capture volumetric information in a single snapshot by encoding both the spatial and angular components of light. Unlike conventional three-dimensional (3D) imaging modalities that rely on mechanical or optical scanning, light-field imaging enables high-speed volumetric acquisition, making it particularly well-suited for capturing rapid biological dynamics. This review outlines the theoretical foundations of light-field imaging and surveys its core implementations across microscopy, mesoscopy, and endoscopy. Special attention is given to the fundamental trade-offs between imaging speed, spatial resolution, and depth of field, as well as recent advances that address these limitations through compressive sensing, deep learning, and meta-optics. By positioning light-field imaging within the broader landscape of biomedical imaging technologies, we highlight its unique strengths, existing challenges, and future potential as a scalable and versatile tool for biological discovery and clinical applications.

## 1. Introduction

Three-dimensional (3D) imaging has emerged as a transformative approach in biomedical sciences, fundamentally reshaping our ability to visualize, analyze, and interpret complex biological systems across scales from cellular processes to whole-organ dynamics. Traditional two-dimensional (2D) microscopy often provides limited spatial context, restricting insights into the intricate relationships within biological structures. In contrast, advancements in 3D imaging technologies, such as confocal, multiphoton, and light-sheet fluorescence microscopy, have revolutionized how researchers capture structural and functional information, providing unprecedented access to the spatial organization, dynamics, and interactions of biological components [1, 2].

To acquire 3D images, most current imaging methods rely on scanning approaches. For example, confocal and multiphoton microscopy typically scan an entire volume point by point, while light-sheet fluorescence microscopy scans the volume plane by plane. This dependence on sequential recording introduces an inherent trade-off between imaging speed and signal-to-noise ratio (SNR): the faster the scan, the shorter the dwell time per pixel or plane, resulting in a lower SNR. Although increasing illumination intensity can mitigate this issue, this strategy is limited when fluorophores approach saturation, a situation frequently encountered in confocal microscopy. Furthermore, elevated illumination intensities can exacerbate photodamage to the sample due to the nonlinear relationship between photodamage and excitation intensity, potentially compromising sample viability.

In contrast, the concept of light-field imaging enables the capture of a complete 3D scene in a single snapshot, entirely eliminating the need for scanning [1,2]. Unlike traditional imaging systems that record only intensity values at each pixel, light-field imaging captures the entire distribution of light rays traveling through space, incorporating both their spatial positions and directional components. This rich, multi-dimensional dataset enables powerful computational capabilities, such as digitally refocusing images at varying depths, performing synthetic aperture imaging, and facilitating high-speed volumetric imaging. By fundamentally changing how optical data is collected and processed, light-field imaging has sparked significant advancements not only in computational photography and computer vision but also in emerging fields such as biomedical sciences.

While the practical realization of light-field imaging has been primarily driven by recent advancements in computational methods and sensor technologies, its theoretical foundations date back to more than a century. Gabriel Lippmann first introduced the concept of integral photography, employing an array of lenslets to simultaneously capture multiple viewpoints of a scene [3]. Herbert Ives subsequently built upon this concept, proposing methods to record and reconstruct 3D images from multi-view image sets [4]. Later, Andrey Gershun introduced the term *light field*, explicitly defining it as the radiance distribution in 3D space [5]. Together, these pioneering contributions laid the theoretical groundwork that has enabled contemporary approaches to computationally reconstruct 3D images from 2D projections.

Since its inception, light-field imaging has gained widespread acceptance across various scientific and technological fields. In this Review, we concentrate specifically on the applications of light-field imaging in biomedical sciences. We begin by outlining the fundamental operating principles underlying this imaging modality, followed by a comprehensive discussion of its key implementations and contributions to the field of biomedical imaging. Finally, we highlight the current challenges in the field and explore promising avenues for future research and technological advancements.

## 2. Principle of light-field imaging

### 2.1 The Plenoptic Function and Dimensionality

Adelson and Bergen introduced the concept of the plenoptic function [6], which became fundamental to the light-field theory. The function characterized the visual information in a scene as a 7D function, encompassing position, direction, wavelength, and time. This plenoptic function can be effectively reduced to a 4D light-field, facilitating manageable data acquisition and reconstruction algorithms for most practical applications involving static, monochromatic, or narrow-band scenarios.

Specifically, the plenoptic function can be expressed in its most general form as $P(x, y, z, \theta, \phi, \lambda, t)$, where $(x, y, z)$ denotes spatial coordinates, $(\theta, \phi)$ represents angular coordinates (directions), $\lambda$ is the wavelength of light, and $t$ indicates time [6,7]. Capturing this entire 7D function in a single snapshot is complex. Consequently, light-field imaging primarily focuses on a subset of these dimensions, typically $(x, y, \theta, \phi)$ for monochromatic or band-limited light for a static scene, resulting in a description as $L(x, y, \theta, \phi)$. Here, $L$ describes the radiance (or intensity) of light traveling through each point at a lateral plane $(x, y)$ in a plane for each angle $(\theta, \phi)$ (**Fig. 1a**).

In many practical systems, the light field is expressed in a two-plane parameterization, $L(u, v, s, t)$. Here, $(u, v)$ corresponds to coordinates on one plane (e.g., an aperture or microlens array), and $(s, t)$ corresponds to coordinates on a second plane (e.g., a camera sensor). This 4D light-field notation encapsulates the essential characteristic that each ray can be uniquely described by its position and direction of travel. By restricting attention to the 4D space, data acquisition and reconstruction become more tractable while preserving crucial angular information (**Fig. 1b**). In practice, rendering the 4D light field simplifies the plenoptic function and parameterizes rays by their intersection with two planes, providing a feasible framework for light-field capture and reconstruction. The arrangement of optical elements ensures that light rays traveling in the same direction end up in the identical or corresponding regions of the camera sensor.

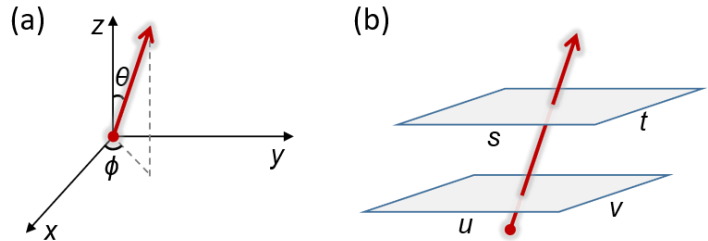

**Figure 1. Diagram of light-field parameterization.** (a) $L(x, y, z, \theta, \varphi) \in R^5$, where $(x, y, z)$ represents the coordinates, and $(\theta, \varphi)$ represents the angles between the light ray and the planes. (b) $L(u, v, s, t) \in R^4$, where $(u, v)$ represents coordinates on one plane and $(s, t)$ on a parallel plane.

### 2.2 Geometric Optics Foundations

Under the assumption of geometric or ray optics, each point on the sensor receives light traveling along a specific trajectory originating from the sample or scene. In conventional imaging, the camera or microscope lens integrates these incoming rays onto a single 2D plane, discarding their angular information. By contrast, in light-field imaging, additional optical elements, such as an array of microlenses or pinholes, partition the incoming ray bundle so that each element forms a miniature image that encodes how these rays are distributed across different directions [6]. This scheme enables the recording of angular details in a single camera snapshot. Using the two-plane parameterization, the raw measurement by the camera sensor (CAM) is $L(u, v, s, t)$. Based on the known geometrical relationship between the position of a ray $r_{CAM}$ on the sensor $q = (s, t)$ and its direction encoded by the optical elements $p = (s-u, t-v)$, the captured image at position $q$ can be modeled as $I(q) = \iint L(u,v,s,t)\,du\,dv = \int_p r_{CAM}(q,p)dp$. Therefore, we can computationally retrieve these rays to reconstruct images by their intersections at different focal depths, thereby achieving effective computational refocusing.

Practical plenoptic implementations typically place the array of microlenses or pinholes in front of an image sensor, using unfocused (*Plenoptic 1.0*, $r_{CAM}(p, q) = r_{MLA}(q\text{-}fp, q/f)$) [8,9], focused (*Plenoptic 2.0*, $r_{CAM}(p, q) = r_{MLA}(-aq/b, -bp/a\text{-}q/f)$) [9,10], defocused ($r_{CAM}(p, q) = r_{MLA}(q\text{-}aq/f\text{-}ap\text{-}bp+abp/f, q/f+p\text{-}bp/f)$) [11], wavefront-coded ($r_{CAM}(p, q) = r_{MLA}(q\text{-}fp, q/f)$) [8], or Fourier configurations ($r_{CAM}(p, q) = r_{MLA}(-qf_{FL}/f_{ML}, q/f_{FL} - pf_{ML}/f_{FL})$) [12,13] (**Fig. 2**). All these light-field renderings enable angular (directional) information and spatial intensity sampling [14,15]. Each aperture or microlens gathers rays from a unique subset of directions passing through that region of space (**Fig. 3**). By recording how the intensity and direction vary, one obtains a dataset that can be subsequently processed for computational refocusing, synthetic aperture imaging, and volumetric reconstruction, all of which are particularly useful in fields such as microscopy, photography, and computer vision [16–20].

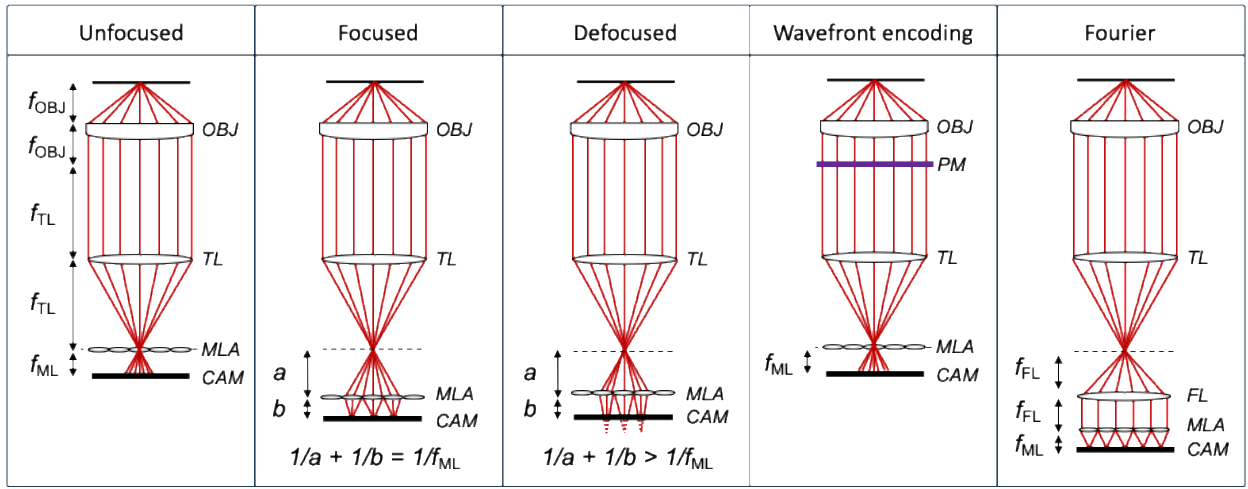

**Figure 2. Configurations of light-field imaging.** Microscope modalities illustrate unfocused (Plenoptic 1.0), focused (Plenoptic 2.0), defocused, wavefront-coded, and Fourier configurations. OBJ, objective lens. TL, tube lens. PM, phase mask. MLA, microlens array. FL, Fourier lens. CAM, camera.

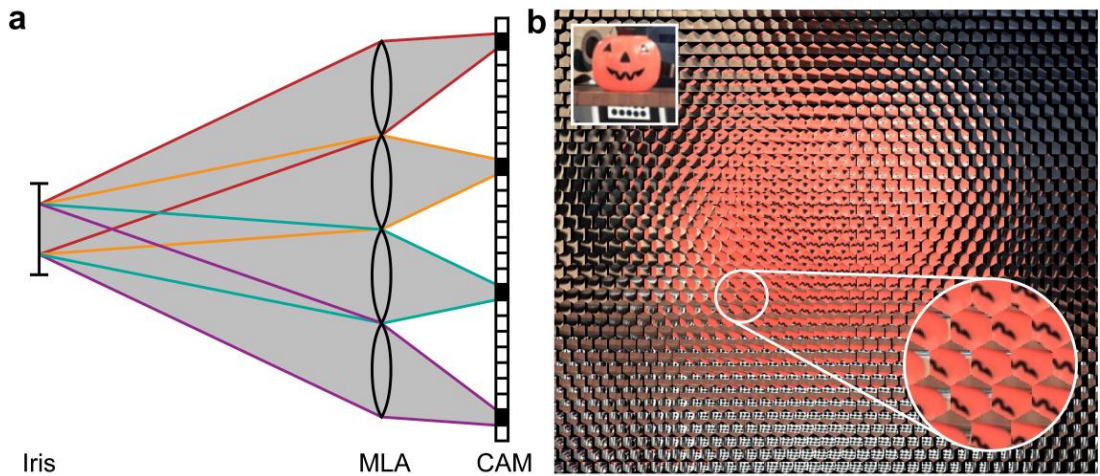

**Figure 3. Schematic of light-field capture using a microlens array.** Rays passing through different parts of the array are recorded in distinct sub-images on the sensor (a), allowing for the computational recovery of angular information (b). MLA: microlens array, CAM: camera sensor. Figures are reprinted with permission from Ref. [16], ACM Press/Addison-Wesley Publishing Co., and Ref. [18], Stanford University.

*2.3 Wave Optics Considerations*

While geometric optics provides a practical and intuitive framework for modeling light propagation in many optical systems, wave phenomena such as diffraction and interference play a significant role in shaping the captured light field and must be accounted for to accurately describe light behaviors. In particular, the point spread function (PSF) of light-field imaging systems may encompass multiple microlenses and apertures, thereby substantially influencing the capability to resolve directional information with high accuracy [21,22]. Modeling these wave effects is therefore essential for achieving high-fidelity reconstructions [23]. This consideration is particularly critical in advanced microscopy, where high-numerical-aperture (NA) objective lenses are engineered to collect light at steep incident angles, thereby allowing for enhanced resolution and image quality [24].

Wave-based light-field models, grounded in the solutions to the Helmholtz or paraxial wave equations, provide more accurate representations of light propagation, interference, and diffraction [25]. Various wave-optics models have been introduced, bridging the gap between geometric and wave optics in light-field imaging [24,26,27] (**Fig. 4**). Specifically, an MLA with a focal length of $f_{ML}$ and a pitch size $d$ can be modeled as $\Phi(\mathbf{x}) = \text{rect}(\mathbf{x}/d)\exp(-ik/(2f_{ML})\|\mathbf{x}\|_2^2)*\text{comb}(\mathbf{x}/d)$, where $\mathbf{x}$ represents the coordinates on the camera sensor plane. Therefore, the light-field PSF $h(\mathbf{x}, \mathbf{p}) = \mathcal{F}^{-1}\{\mathcal{F}\{\Phi(\mathbf{x})U_i(\mathbf{x}, \mathbf{p})\}\exp[-i\lambda/(4\pi)f_{ML}(\omega_x^2+\omega_y^2)]$, where $\mathcal{F}\{\bullet\}$ is the Fourier transform operator, $U_i(\mathbf{x}, \mathbf{p})$ is the light field on the native image plane, $\omega_x$ and $\omega_y$ are spatial frequencies in x and y directions, $\lambda$ is the light wavelength, and $\mathbf{p}(p1, p2, p3)$ represents the coordinates of a point source. Notably, wave-based light-field representations are, by nature, relatively more sophisticated and computationally demanding. However, they yield higher accuracy, especially in biological microscopy and other high-resolution imaging scenarios, as well as when reconstructing complex or scattering samples [28].

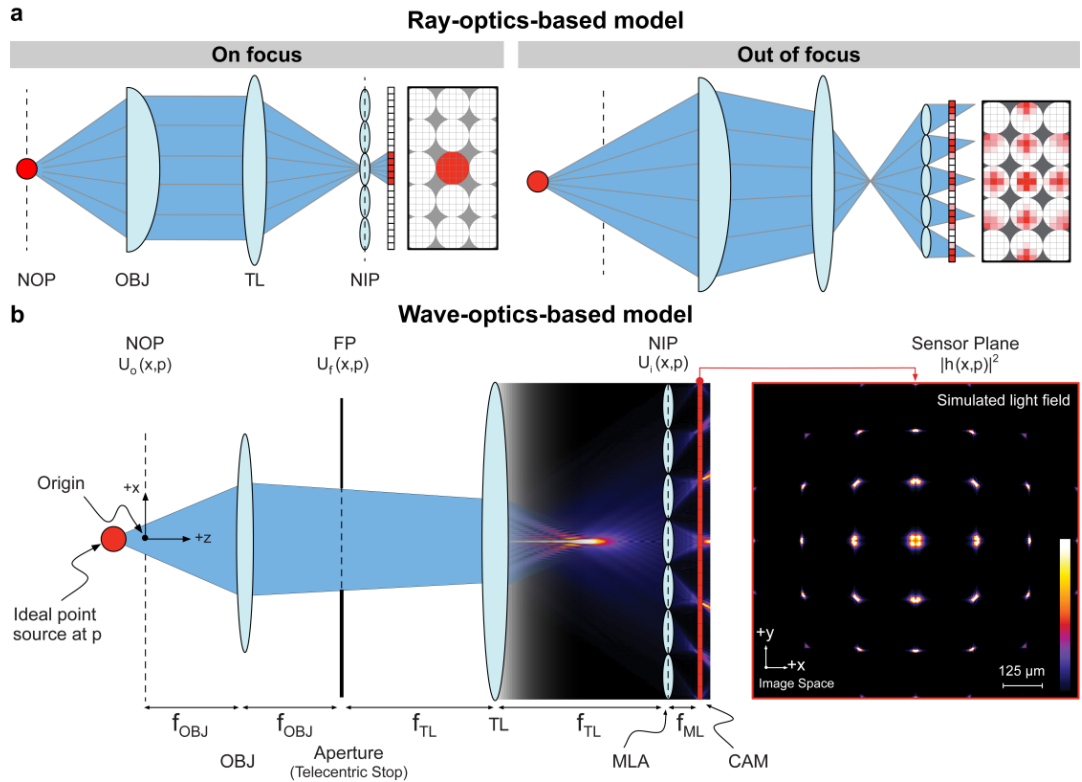

**Figure 4. Comparison of ray-optics-based (a) versus wave-optics-based (b) modeling in high-NA systems.** Figures are reprinted with permission from Ref. [24], Optica (OSA).

*2.4 Light-Field Rendering and Visualization*

A crucial advantage of light-field imaging lies in its ability to render or synthesize images from multiple viewpoints or focal planes without requiring additional physical acquisitions. Light-field rendering presents four prominent features. *(1)* Refocusing and focal stack generation [16,18,29]: computationally shifting and integrating the sub-images captured by each microlens allows for "refocusing" the scene at different depths. This process yields a stack of images, each focused on a particular plane, enhancing 3D visualization and analysis. *(2)* Synthetic aperture rendering [30]: captured light fields can simulate a larger synthetic aperture by coherently combining rays from different directions. This approach reduces occlusions and can yield shallow depth-of-field images, which are useful for photography, product inspection, or artistic rendering. *(3)* Multi-view or stereoscopic outputs [31]: because each microlens captures a different perspective, the recorded data can be projected into multiple viewing angles, supporting stereoscopic or multi-view displays. *(4)* Ray-tracing and volumetric reconstruction: advanced rendering pipelines leverage tomographic or wave-optics-based inversion to reconstruct volumetric datasets [24], particularly valuable in microscopy. These rendering and visualization capabilities are central to light-field technology in various fields, including cinematography, consumer photography, microscopy, and medical imaging.

## 3. Implementations and Applications in Biomedicine

Light-field imaging has rapidly gained momentum in biomedical research, enabling the acquisition of single-shot volumetric data, a particularly advantageous approach for investigating dynamic biological processes. From single-cell analyses and whole-organism studies to functional imaging, light-field systems offer a synergistic combination of high-speed 3D capture, seamless integration with conventional microscopy platforms, and reduced phototoxicity compared to traditional scanning-based methods. This section provides an overview of major light-field implementations in biomedical settings, highlighting representative applications that illustrate the growing influence of light-field imaging in the biomedical sciences.

*3.1 Light-field Microscopy*

Fluorescence microscopy has long been a cornerstone of biomedical research. It offers molecular specificity, high contrast, and robust labeling methods for visualizing specimens at subcellular resolution [32,33]. Conventionally, major fluorescence microscopy techniques produce orthographic views and acquire 3D information in a sequential or scanning fashion [34,35]. Although these techniques enable optical sectioning and deep tissue imaging, their reliance on point-by-point or plane-by-plane acquisition inherently lowers temporal resolution. Additionally, since scanning modalities tightly focus the laser into a small spot, the elevated light fluence significantly increases the risk of photodamage, posing a major challenge for intravital and long-term studies. In contrast, emerging light-field microscopy (LFM) techniques capture the 2D spatial and 2D angular light distributions in a single camera exposure [36,37]. By leveraging computational reconstruction algorithms, this 4D dataset can be refocused or reconstructed into a volumetric view of the specimen without mechanical scanning and achieving enhanced spatiotemporal performance [24,36–38] (**Fig. 5**). Mathematically and computationally, they can be achieved by solving the inverse problem of $\mathbf{f} = \mathbf{Hg}$, where $\mathbf{f}$ is the measured light-field images, $\mathbf{H}$ is the discrete PSF, and $\mathbf{g}$ is the corresponding volume. Using the Richarson-Lucy algorithm, the volume can be iteratively estimated with $\mathbf{g}^{(k+1)} = \text{diag}(\mathbf{H}^T\mathbf{1})^{-1}\text{diag}(\mathbf{H}^T\text{diag}(\mathbf{Hg}^{(k)})^{-1}\mathbf{f})\mathbf{g}^{(k)}$. Such an approach significantly improves volumetric acquisition speed (ultimately limited by the camera frame rate) while minimizing photobleaching and phototoxic effects, thereby enhancing the feasibility of time-lapse imaging (**Table 1**).

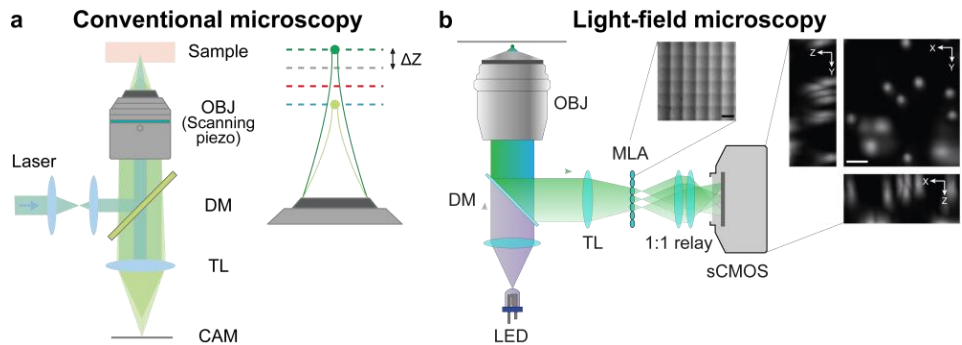

**Figure 5. Comparison of conventional (a) and light-field microscopy (b).** Figures are reprinted with permission from Ref. [34], Optica (OSA), and Ref. [38], Springer Nature.

These advantages of LFM have proven especially transformative in functional brain imaging, where rapid volumetric data acquisition is critical for capturing millisecond-scale neuronal and synaptic activity [24,38]. LFM has achieved cellular resolution through tissue depths of tens to hundreds of micrometers, validated using diverse model systems, including C. elegans, Drosophila, zebrafish, and mice [24,37–40]. This enables large-scale, high-speed recordings of network dynamics. Various algorithms have been developed to enhance light-field image processing and visualization [28,41–44]. Meanwhile, at the microscopic scale, LFM has demonstrated promising results in high-resolution 3D imaging of single-cell specimens, facilitating the study of subcellular organelles and rapid intracellular processes [11]. Additionally, incorporating scanning mechanisms and adaptive optics has extended the utility of LFM to more complex, multicellular samples [45–48]. By mitigating light scattering and aberrations inherent in thicker tissues, these hybrid approaches further expand the range of biological questions addressable by LFM, underscoring its potential as a powerful tool for multi-scale, time-resolved fluorescence imaging (**Fig. 6**).

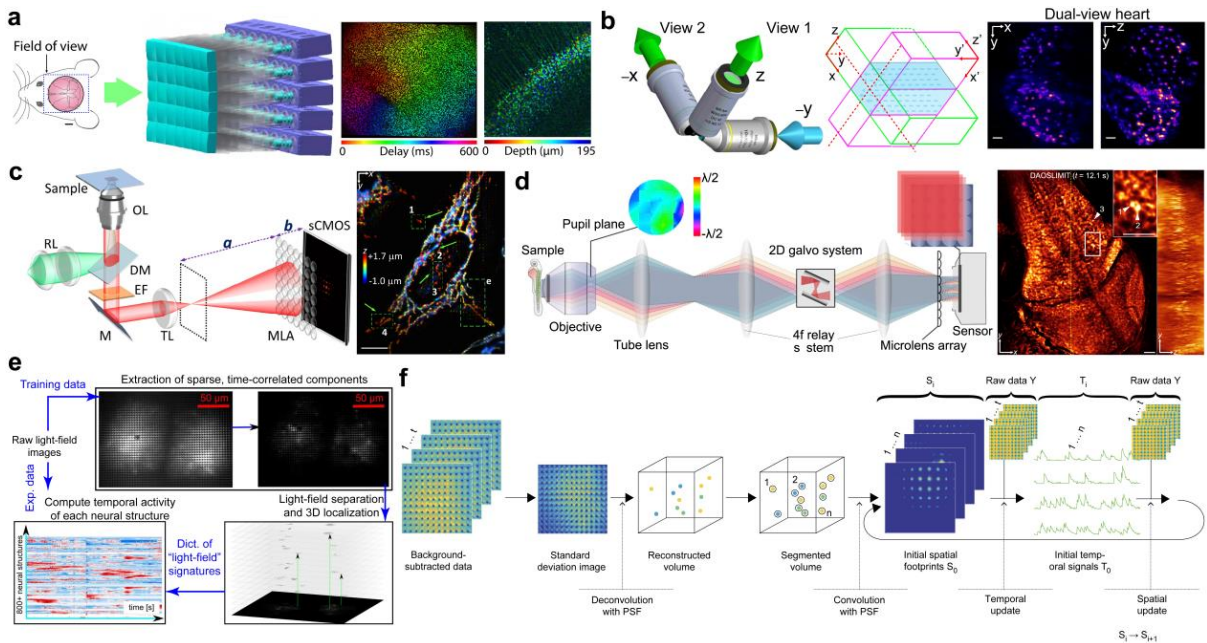

**Figure 6. Summary of current LFM techniques and performance.** (a) Real-time, ultra-large-scale, high-resolution (RUSH) imaging platform for brain-wide structural-functional investigation of awake, behavioral mice. (b) Selective-volume illumination with simultaneous acquisition of orthogonal light fields of a beating heart. (c) Fast, volumetric live-cell imaging using high-resolution light-field microscopy. (d) The sLFM system for high-resolution spatial and angular measurements. (e) Compressive light-field image processing for samples tagged with engineered fluorescent proteins to track brain activity. (f) Key steps of seeded iterative demixing (SID) microscopy. Figures are reprinted with permission from Refs. [39,40,42], Springer Nature, Refs. [11,41], Optica (OSA), Ref. [45], Cell Press.

More recently, Fourier light-field microscopy (FLFM), also known as extended light-field microscopy (XLFM), has emerged as a pivotal advancement, offering an alternative solution to traditional light-field architectures by sampling spatial frequencies rather than solely the spatial domain [12,13,49,50]. This approach emerged from the need to overcome the intrinsic non-uniform sampling of conventional microlens-based methods, which often generate reconstruction artifacts, limit volumetric resolution, and increase computational overhead [50–53]. The demonstrations have shown improved imaging capabilities and extended applications in neuroscience, cell biology, and organoid research [49,54–60]. Today, FLFM systems continue to expand across a wide range of platforms and biological model systems, paving the way for rapid, high-throughput studies of cellular and tissue-level functions [61–69] (**Fig. 7**).

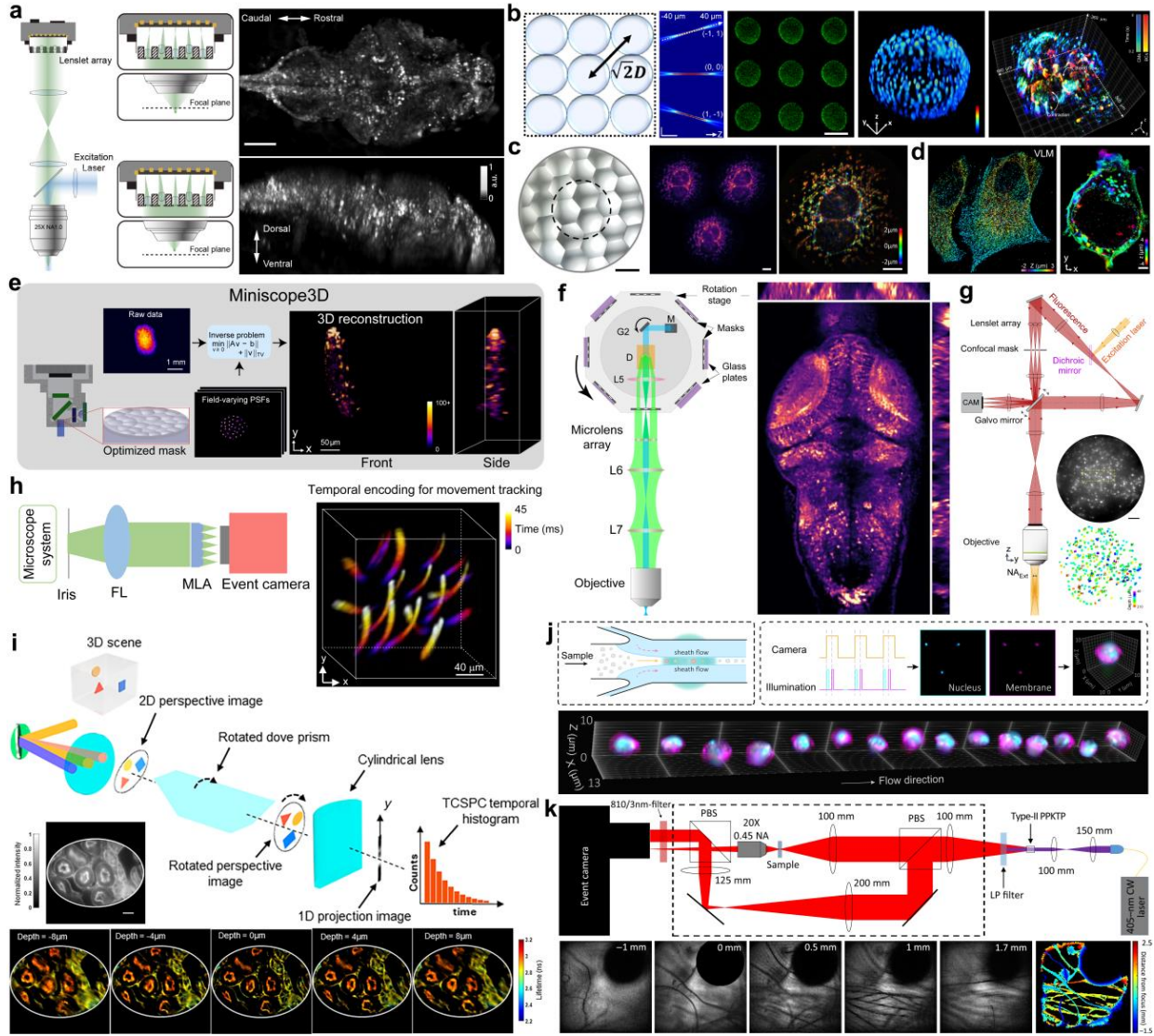

**Figure 7.** Summary of current LFM techniques and performance. (a) Whole brain imaging of larval zebrafish with XLFM. (b) Observation of human colon organoids and the embryonic heart of tadpoles using hPSF-FLFM. (c) Imaging mitochondria in fixed mammalian cells using HR-FLFM. (d) Super-resolved images of HeLa cells and Jurkat T cells captured with single-molecule light-field microscopy. (e) Fluorescent bead imaging using the Miniscope 3D system. (f) Whole-brain functional imaging of neural activity in movement-restrained larval zebrafish using confocal LFM. (g) Volumetric voltage imaging and 3D labeling of neuronal populations in the awake mouse cortex using confocal LFM. (h) Time color-coded 3D motion trajectory reconstructed over a 45 ms time span using EventFLFM. (i) Imaging of a mouse kidney tissue section using light-field tomographic FLIM (LIFT-FLIM). (j) Imaging of membrane- and nucleus-labeled human activated T cells in flow using light-field flow cytometry (LFC). (k) Experimental setup of quantum correlation LFM (QCLFM) and refocusing of lens tissue fibers. Figures are reprinted with permission from Refs. [52,54,58,63,65], Springer Nature, Refs. [50,55,56], Optica (OSA), Ref. [49], Elife, Ref. [57], Elsevier, Ref. [61], Proceedings of the National Academy of Sciences, and Ref. [62], American Physical Society, Ref. [59], bioRxiv.

**Table 1. Summary and comparison of current light-field microscopy techniques.**

| | Ref. | Light-field acquisition methods | Lateral resolution (μm) | Axial resolution (μm) | FOV (μm) | DOF (μm) | Imaging speed (Hz) | Experimental Demonstration |
|---|---|---|---|---|---|---|---|---|
| Current LFM techniques | [37] | Micro lens array | 6.9 | 18.4 | 15000 × 15000 | 139 | N.A. | Live skin patch from longfin inshore squid *Loligo pealeii* |
| | [38] | Micro lens array | 1.36 | 2.55 | 700 × 700 | 200 | 5-50 | Whole-brain $Ca^{2+}$ imaging in C. elegans and larval zebrafish at single-neuron resolution |
| | [41] | Microlens array + Compressive Sensing | 4 | N.A. | 200 × 200 | 40-200 | 100 | Brain activity tracking in live zebrafish |
| | [42] | Microlens array + seeded iterative demixing (SID) | 2-3 | 5-10 | 900 × 900 | 380 | 30-100 | Source localization in zebrafish larvae, volumetric $Ca^{2+}$ imaging in mouse brain |
| | [39] | Relay lens array + sCMOS array | 1.2 | 1.2 | 10000 × 12000 | 150-200 | 30 | Brain-wide structural imaging and functional imaging in awake, behaving mice |
| | [40] | Dual-view Microlens Array | 1.2-2.8 | 1.7-4.1 | 300 × 300 | 300 | 200 | Beating heart and blood flow imaging in juvenile medaka fish at single-cell resolution |
| | [11] | Microlens array | 0.3-0.7 | 0.3-0.7 | 133 × 133 | >3 | 10-1000 | Fixed HeLa cells, live COS-7 cells, and live drp1-/- mouse embryo fibroblasts |
| | [45] | Scanning LFM + digital adaptive optics | 0.22 | 0.4 | 225 × 225 | 16 | 50-500 | 3D intravital subcellular imaging in zebrafish, mice, and mammals |
| | [46] | Virtual Scanning LFM + Deep Learning | 0.23 | 0.42 | 210 × 210 | 18 | 500 | Robust high-resolution snapshot 3D imaging in mammals, and 3D voltage imaging in Drosophila |
| Current FLFM/XLFM techniques | [49] | Microlens array | 3.4 | 5.0 | 800 × 800 | 400 | 50-100 | Whole brain imaging of larval zebrafish |
| | [13] | Microlens array | 3.9-6.2 | N.A. | 1386 × 1386 | 77-240 | N.A. | Air force target and cotton fiber |
| | [50] | Microlens array | 2.12 | 4.70 | 67 × 67 | 64 | N.A. | Pollen grains and mouse kidney |
| | [51] | Microlens array + dark field | 12 | N.A. | 2300 × 2300 | 1050 | N.A. | Bubble samples and zebrafish |
| | [52] | Miniaturize + microlens array | 2.76 | 15 | 900 × 700 | 390 | 40 | GFP-tagged neurons in the mouse brain and two different samples of freely moving tardigrades |
| | [56] | Microlens array + single molecule localization | 0.02 | 0.02 | 15 × 15 | 6 | N.A. | Membranes of fixed eukaryotic cells and DNA nanostructures |

| | [54] | Confocal + microlens array | 2 | 2.5 | 800 × 800 | 200 | 70 | Neural activities over the whole larval zebrafish brain and scattering mouse brain |
|---|---|---|---|---|---|---|---|---|
| | [55] | Microlens array + hybrid PSF | 0.3-0.7 | 500-1500 | 70 × 70 | 4 | 200 | Mitochondria and peroxisomes in COS-7 cells |
| | [57] | Microlens array + hybrid PSF | 2-3 | 5-6 | 900 × 900 | 200 | 100 | Cellular dynamic processes of whole organoids in response to rapid extracellular physical cues |
| | [58] | Microlens array + event camera | 3.9 | 21.0 | 130 × 130 | 300 | 1000 | Blinking neuronal signals in scattering mouse brain tissues and 3D tracking of GFP-labeled neurons in freely moving C. elegans |
| | [61] | Dove prism + cylindrical lens array + tomography + lifetime imaging | 1.8 | 3.0 | 227 × 143 | 16 | 0.056 | Mouse kidney tissue section, human lung cancer pathology slide, and lung organoids |
| | [62] | Event camera + quantum correlation | 10 | 30-40 | N.A. | 2000 | 0.001-0.01 | Air force target and lens tissue fiber |
| | [63] | Microlens array + confocal | 4 | 12 | 800 × 800 | 180 | 400 | Volumetric voltage imaging of neuronal populations in the awake mouse brain |
| | [65] | Microlens array + microfluidics + stroboscopic illumination | 0.4-0.6 | 0.4-0.6 | 70 × 70 | 3-6 | 200 | Flowing HeLa cells, mouse blood cells, mouse and human T cells, platelets, apoptosis in Jurkat cells, and RNA delivery in mouse spleen cells and endothelial cells |

*3.2 Light-field Mesoscopy*

Optical mesoscale imaging, or mesoscopy, refers to the imaging of objects that fall between microscopic and macroscopic scales, typically spanning dimensions from millimeters to centimeters, while still maintaining cellular resolution [70]. This field bridges the gap between traditional microscopic biological imaging and macroscopic clinical imaging, connecting detailed cellular-level insights with clinically relevant larger-scale observations. With the growing use of 3D cell/tissue cultures [71–73], intact organs [74–76], and whole-animal models [77] in life sciences, mesoscopic imaging has become increasingly important. It facilitates the investigation of more complex organ structures and enables observation of later developmental stages that were previously inaccessible or challenging to visualize.

Mesoscopic imaging has been explored through various techniques, including Mesolens [78], two-photon mesoscope [79,80], mesoscale selective plane illumination microscopy (MesoSPIM) [81], high-resolution (RUSH) mesoscope [39], and optical projection tomography [82]. The development of these mesoscopic imaging systems is challenged by scale-dependent geometric aberrations inherent to optical elements [83]. This results in a trade-off between the achievable space-bandwidth product (SBP) and the complexity of the optical design [83,84], as seen in mesoscopes using sequential [78–80] and multiscale lens designs [39]. Moreover, scanning-based mesoscopic imaging systems face fundamental limitations imposed by the scanning speed of mechanical components, restricting their capability for large-scale, high-resolution volumetric imaging.

To overcome these challenges, computational imaging approaches such as light-field microscopy are actively being explored (**Fig. 8**). For example, Nöbauer et al. introduce a mesoscale light-field (MesoLF) system for high-speed volumetric functional imaging of the mouse cortex, capable of simultaneously recording activity from over 10,000 active neurons across a volume of $\phi 4 \times 0.2\ mm$ at 18 volumes per second [85]. The MesoLF utilizes a scalable computational pipeline for neuronal localization and signal extraction. This pipeline incorporates a phase-space-based deconvolution method, accompanied by a background 'peeling' procedure, which effectively reduces reconstruction artifacts and significantly enhances reconstruction quality. However, the current achievable imaging depth is limited to ~400 μm primarily due to the loss of directional information caused by photon scattering. Additionally, the current imaging speed of 18 volumes per second, limited by photon detection efficiency and camera read noise, could potentially be improved with future advancements in camera technology.

Notably, recent advancements in light-field mesoscopy are trending toward miniaturization, making it increasingly practical for imaging freely moving animals. For instance, Xue et al. present a computational miniature mesoscope (CM$^2$) that achieves single-shot 3D wide-field fluorescence imaging with a large field of view $8.1 \times 7.3 mm$ and a $2.5 mm$ depth of field. The CM$^2$ integrates a microlens array for imaging and a light-emitting diode array for excitation in a compact design. The entire system measures approximately $29 \times 29 \times 30$ mm³ and weighs only 19 grams. However, due to the small disparity baseline, the system initially exhibited relatively low axial resolution (~200 μm) [53]. The authors later introduced an updated version, CM$^2$ V2, which incorporates a hybrid emission filter for improved image contrast and a free-form collimator for enhanced illumination efficiency [86]. Additionally, they reduced the system footprint to approximately $36 \times 36 \times 15$ mm³. CM² V2 also employs a deep-learning-based reconstruction algorithm, CM2Net, trained with a 3D linear shift-variant imaging formation model, enabling faster, higher-resolution reconstructions and significantly improving the axial resolution to approximately 25 μm (**Table. 2**).

Light-field mesoscopy enables bioimaging with a large field of view; however, this advantage comes with the trade-off of increased optical aberrations at the image periphery [85,86]. Combining light-field imaging with digital adaptive optics has proven effective in addressing this problem. For example, Zhang et al. introduced a compact real-time, ultra-large-scale, high-resolution 3D mesoscope (RUSH3D) that utilizes a scanning light-field framework. This system achieves uniform resolutions of $2.6 \times 2.6 \times 6$ μm over a volume of $8000 \times 6000 \times 400$ μm³ at a frame rate of 20 Hz [87]. Additionally, the system integrates digital adaptive optics and multiscale background rejection (MBR) to enhance imaging quality and reduce aberrations in turbid tissues. Although its complexity and associated costs remain notable limitations, the system delivers orders-of-magnitude improvements in data throughput, system size, and overall efficiency compared to conventional imaging methods.

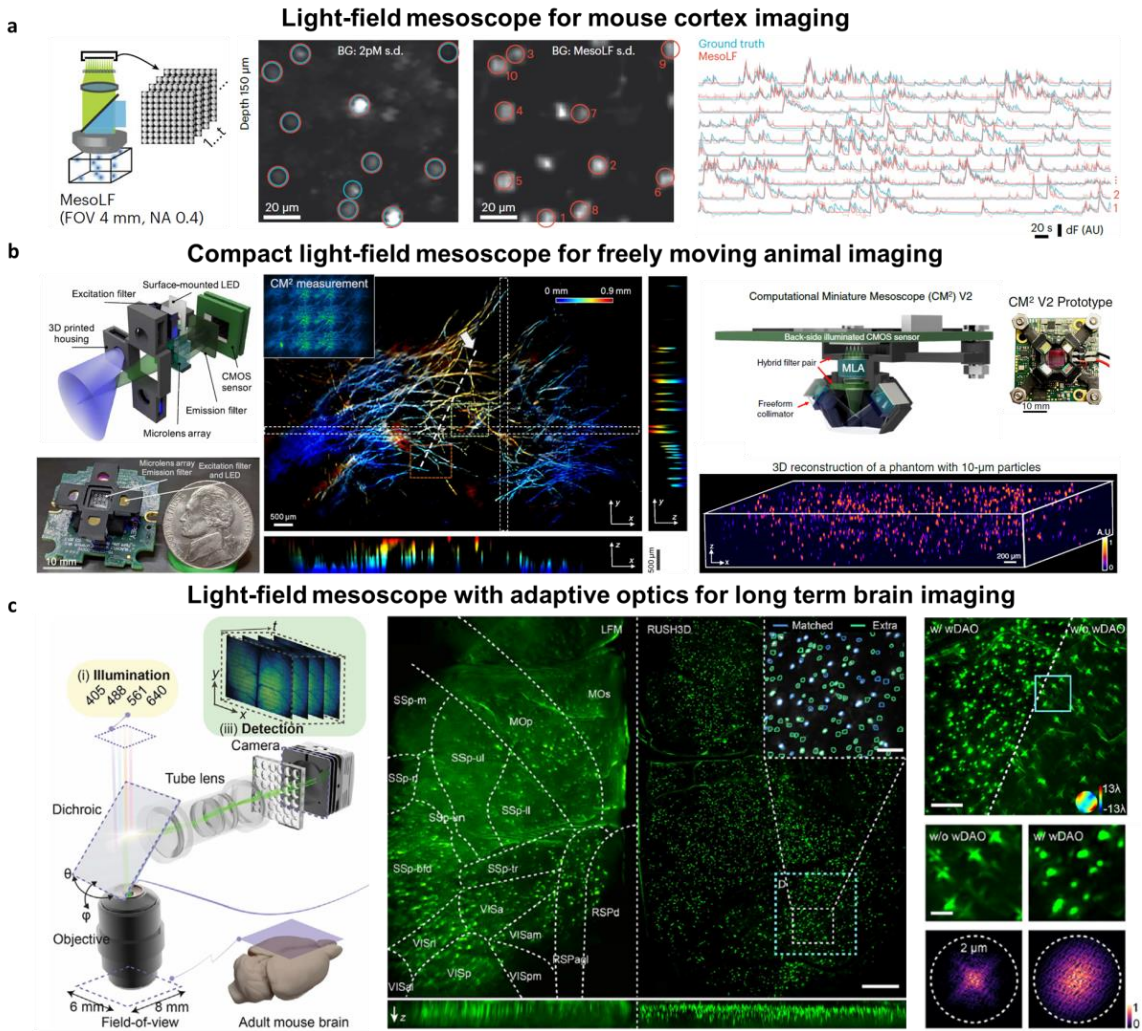

**Figure 8.** Light-field mesoscope. (a) Mouse cortex imaging with conventional light-field detection. (b) Computational miniature mesoscope that achieves single-shot 3D wide-field fluorescence imaging. (c) large-scale 3D recording of calcium activities at single-neuron resolution with an adaptive-optics-enhanced light-field mesoscope. Figures are reprinted with permission from Refs. [85,87], Springer Nature, Ref. [53], AAAS Science, Ref. [86], OPTICA

**Table 2. Summary and comparison of current light-field mesoscope modalities.**

|  | FOV(mm) | DOF(mm) | Lateral resolution(um) | Axial resolution(um) | Imaging speed(vps) | Experimental demonstration |
|---|---|---|---|---|---|---|
| MesoLF [85] | 4 × 4 | 0.2 | 7 | 32 | 18 | Neural activity (Ca+ imaging) across large cortical areas in mice |
| CM² [53] | 8.1 × 7.3 | 2.5 | 7 | 200 | N.A. | Fluorescent tissue and fibers |
| CM² V2 [86] | 7 × 7 | 0.8 | 6 | 25 | N.A. | Mixed fluorescent beads |
| RUSH3D [87] | 8 × 6 | 0.4 | 2.6 | 6 | 20 | Neural activity (Ca+ imaging) across large cortical areas in mice |

*3.3 Light-field Endoscopy*

Endoscopy is a critical, minimally invasive tool for diagnosing and treating internal conditions through natural openings or small incisions. It has significantly transformed modern medicine by reducing infection risks, shortening patient recovery time, and lowering medical costs [88]. However, conventional endoscopic systems primarily provide 2D projections of 3D anatomical structures, limiting depth perception and spatial understanding. This limitation is particularly critical in complex procedures such as tumor resections or vascular surgeries, where precise 3D spatial information is essential for accuracy and safety [89–91].

To address these limitations, several advanced optical approaches for 3D endoscopic imaging have been developed, including methods using stereoscopic optics [92–94], all-fiber-optics-based scanning [95,96], liquid crystal or electrowetting lenses [97–99], or holography [100]. Among these, light-field imaging has emerged as a promising solution due to its scalability and volumetric imaging capabilities. This approach offers several key advantages: real-time 3D imaging for informed decision-making, compact and scalable designs suitable for integration into off-the-shelf endoscopes, and quantitative depth measurement, which is critical for robotic surgeries and precise diagnostics. Furthermore, light-field imaging reduces operator dependence, addressing challenges such as dizziness associated with stereoscopic systems [101]. These attributes position light-field imaging as a transformative technology for overcoming the limitations of traditional endoscopy (**Fig. 9**).

There are two primary approaches to integrating light-field imaging into endoscopic systems: incorporating a light-field module downstream of traditional endoscopic optics or replacing the original entrance optics with microlens arrays (**Table 3**). The integration of light-field imaging technology into medical endoscopes has been demonstrated in several studies [102–104]. These systems use a microlens array placed at the image plane of a conventional endoscope to capture 4D light-field data in a single shot. This allows for reconstruction of 3D images and digital refocusing, thereby overcoming the limitations of traditional 2D endoscopic imaging. For example, Liu et al. introduced light-field imaging into an off-the-shelf endoscope to enable 3D imaging without modifying the original probe system, facilitating its ready adoption for clinical use [103].

However, due to the small numerical aperture of most commercial endoscopic optics, directly integrating light-field imaging with off-the-shelf endoscopes often results in inferior depth resolution. To address this limitation, Bedard et al. developed a light-field otoscope from the ground up. By designing the light-field imaging module and front-end optics in a synergistic manner, they achieved 3D imaging of the tympanic membrane with sub-millimeter depth accuracy while maintaining a compact device size suitable for pediatric use [102]. In another example, Zhu et al. introduced a light-field laryngoscope for 3D imaging of the vocal folds, utilizing a custom combination of an objective lens and a gradient refractive index(GRIN) lens to relay images, achieving a depth resolution of 0.37 mm [104]. Additionally, they incorporated a high-resolution reference imaging channel into their probe and implemented an image fusion algorithm to enhance lateral spatial resolution, addressing the intrinsic limitation of light-field imaging in achieving diffraction-limited lateral resolution.

Rather than positioning the light-field imaging module at the proximal end of the endoscopy probe, it can be directly integrated into the distal end, replacing the original entrance optics. For example, Hassanfiroozi et al. developed a light-field endoscope incorporating a hexagonal array of liquid-crystal (LC) lenses with convex-ring electrodes, enabling 3D imaging with a tunable depth range [98]. The authors later improved this design by introducing a multifunctional LC lens (MFLC-lens) featuring a dual-layer electrode structure, allowing for electronic switching between 2D and 3D imaging modes as well as focal adjustment [99]. Lee et al. demonstrated a similar design, utilizing an electrowetting lens array as the light-field imaging module, which enables fast focal length adjustment and 2D/3D switching. They also highlighted its advantages over conventional lens arrays with fixed focal lengths [90].

Despite its compact size, the clear aperture of each lenslet and the baseline disparity in the array are fundamentally constrained by packing density and the overall dimensions of the probe's distal end, leading to reduced resolution. To address this limitation, Wang et al. introduced a 3D integral-imaging endoscope that employs a single, large-aperture, tunable liquid crystal lens at the distal end of the probe [97]. Instead of capturing individual perspective images in parallel, their system translates the probe to acquire multiple viewpoint images, mimicking synthetic aperture integral imaging. This approach achieves both high spatial resolution and an extended depth of field, overcoming the trade-offs associated with conventional light-field endoscopic designs.

Besides microlens arrays, recent advances in light-field endoscopy have also explored the use of GRIN lenses to demultiplex light rays. Due to their compact size compared to conventional compound lenses, GRIN lenses enable practical array integration at the entrance optics of an endoscope while effectively capturing angular information necessary for 3D reconstruction. For example, Guo et al. proposed a light-field micro-endoscopy system incorporating a GRIN lens array [89]. Their system achieves spatial resolutions of 20–60 μm laterally and 100–200 μm axially across an imaging volume of approximately 5 mm × 5 mm × 10 mm. More recently, the authors optimized their optical design and integrated the GRIN lens array with a rigid endoscope [101]. This improved system achieves a 3D resolution of approximately 100 μm over a depth range of ~22 mm, with a lateral field of view extending up to 1 cm².

So far, most light-field endoscopy techniques have been implemented in rigid endoscopes. Compared to rigid endoscopes, flexible endoscopes using fiber bundles offer several advantages, including a lower complication rate, improved patient comfort, the ability to perform procedures without general anesthesia, and access to tissues that are typically unreachable with rigid endoscopes, such as those in the gastrointestinal tract and pulmonary airways. However, integrating light-field imaging with flexible endoscopy presents significant challenges, as the angular information of the light-field is generally considered lost or scrambled when transmitted through flexible fiber bundles.

To address this challenge, Zhou et al. positioned a microlens array at the distal end in front of a fiber bundle, transmitting individual perspective images, rather than the light field, through the bundle [91]. They demonstrated 3D endoscopic imaging within a 333-μm-diameter field of view, achieving a depth of field of 24 μm and spatial resolution of up to 3.91 μm near the focal plane. However, fiber bundles are fundamentally constrained by their pixel count, bundle diameter, and flexibility requirements. For instance, current small-diameter image guides typically provide pixel counts ranging from a few thousand to tens of thousands, which is significantly lower than the megapixel resolutions generally required for high-quality light-field imaging. This limitation hampers the capability to transmit high-resolution perspective images through these systems.

Alternatively, Orth et al. demonstrated that standard optical fiber bundles can be directly used for light-field sensing by leveraging the modal structure within the fiber cores to extract depth information [105]. This method enables digital refocusing, stereo visualization, and surface and depth mapping. The key insight behind their approach is that the angular dimension of the light field is encoded within the intracore intensity patterns of the fiber bundle, which have traditionally been overlooked. By quantitatively relating these intensity patterns—arising from angle-dependent modal coupling—to the angular structure of the light field, they showed that optical fiber bundles can achieve single-shot surface and depth mapping with an accuracy better than 10 μm at distances of up to ~80 μm from the fiber bundle facet. Additionally, this approach is resilient to fiber bending, making it well-suited for clinical translation. However, a limitation of this method is that it requires the sample to be sparse and non-overlapping along the depth axis, and it suffers from a short working distance, restricting its applicability in imaging complex tissue structures.

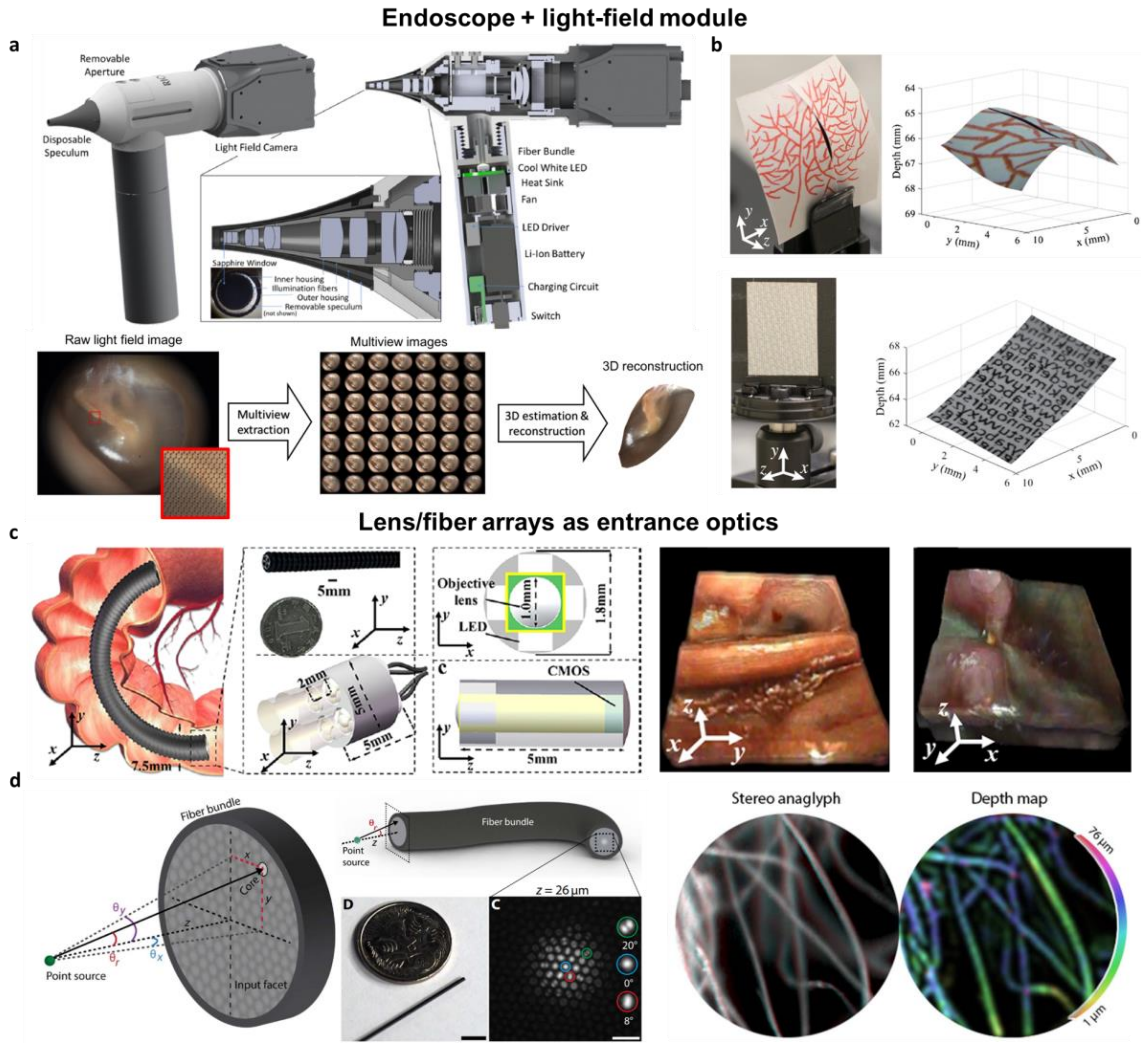

**Figure 9.** Light-field endoscope. (a) light-field otoscope for in-vivo tympanic membrane. (b) Snapshot light-field laryngoscope 3D vocal phantom imaging(c) Fiber bundled based light-field endoscope for stomach and rectum imaging of a rabbit. (d) Optical fiber bundle micro-endoscopes. Red-cyan stereo anaglyph of lens paper tissue with highlighter ink. Figures are reprinted with permission from Ref. [102,106], OPTICA, Ref. [105], AAAS Science, Ref. [104], SPIE.

**Table 3. Summary and comparison of current light-field endoscope modalities.**

| | Ref. | Probe dimension (mm) | Light-field acquisition methods | Mono-chromatic/chromatic | Lateral resolution (um) | Axial resolution(um) | FOV (mm) | DOF (mm) | Experimental Demonstration |
|---|---|---|---|---|---|---|---|---|---|
| Traditional endoscope + light-field module | [102] | N.A. | Micro lens array | chromatic | 56.8 | 60 | 15 | 4 | Human tympanic membranes in normal and otitis media conditions |
| | [104] | Ø10 | Micro lens array | monochromatic | 36.6 | 370 | N.A. | 5 | Vocal fold phantom |
| Lens or fiber | [99] | N.A. | Tunable liquid crystal (LC) lens | chromatic | N.A. | N.A. | N.A. | 12 - 51 | A ladybug, a piece of paper printed with "23," and a butterfly used as a 3D scene |

| [89] | N.A. | GRIN lens array | mono-chromatic | 20–60 | 100–200 | 5*5*10 | 10 | A piece of mandarin orange albedo, a papilla of bovine omasum, and a fingertip |
| [107] | N.A. | Fiber bundles | mono-chromatic | N.A. | N.A. | N.A. | 0.1 | Proflavine-stained brain slice, skin surface autofluorescence |
| [101] | 52.5*5*5 | GRIN lens array | chromatic | ~100 | ~800 | 10*10*22 | 22 | Phantom target resembling red and blue blood vessels / phantom target of heart |
| [106] | 5*5*5 | Multi-camera array | mono-chromatic | 31 | 255 | 2.3*2.3*10 | 10 | Rabbit's digestive tract. |
| [91] | 500*2*2 | Fiber bundles | mono-chromatic | 3.91 | 35.4 | 0.33 | 0.024 | Human skin tissue section and HeLa cells |

## 4. Discussion and outlook

### 4.1 Speed limit

A primary advantage of light-field imaging lies in its high-speed volumetric acquisition capability. Because volumetric data can be acquired simultaneously rather than sequentially, the 3D imaging speed of light-field systems is primarily limited by the camera's readout rate. However, capturing the complete four-dimensional light-field dataset ($x, y, \theta, \phi$) produces a significant amount of data, requiring large-format image sensors to fully sample the incoming light-field. Unfortunately, these high-resolution sensors typically operate at reduced frame rates, resulting in an inherent trade-off between imaging speed and spatial resolution in practical implementations of light-field imaging systems.

Addressing the "big data" challenge in light-field imaging requires innovative sampling strategies. One crucial insight is that the primary objective of biomedical light-field imaging is often the generation of volumetric images, rather than capturing the complete spatial and angular information of every individual light ray. Thus, it is unnecessary to measure every voxel within the full four-dimensional ($x, y, \theta, \phi$) dataset. Instead, employing compressive sensing techniques can significantly enhance acquisition efficiency, provided the sample exhibits sparsity in certain domains.

One effective strategy is to leverage the temporal sparsity inherent in biological events to compress the acquired light-field data. For example, Guo et al. developed a neuromorphic light-field imaging approach that replaces conventional 2D intensity cameras with event-based cameras. Rather than recording complete spatiotemporal datasets, this technique captures only pixel-intensity changes, significantly reducing data volume and optimizing bandwidth usage. As a result, their system achieves imaging speeds at kilohertz frame rates. The authors demonstrated their method by imaging simulated neuronal signals within scattering mouse brain tissue and tracking green fluorescent protein (GFP) labeled neurons in freely moving C. elegans in 3D. However, the spatial resolution is currently limited due to the low pixel count of event-based sensors. Additionally, the reduced sensitivity and binary output format of event cameras restrict their ability to detect subtle neuronal signals, such as subthreshold voltage oscillations.

In another approach aimed at compressing data in the spatial domain, Wang et al. introduced a squeezed light-field microscopy (SLIM) method [108]. This method leverages the observation that perspective images captured by conventional light-field cameras contain highly redundant spatial information, with differences primarily arising from depth-dependent disparities between sub-aperture views. SLIM exploits this redundancy by compressing perspective images along different orientations and redistributing these compressed measurements across multiple views. Such spatial compression drastically reduces the size of the dataset, enabling high-resolution 3D imaging with a single, low-pixel-count camera that supports high-speed readout. Wang et al. demonstrated the effectiveness of SLIM through kilohertz-scale volumetric imaging of diverse biological systems, including in vivo detection of neural action potentials and subthreshold oscillations in mice.

Although the substantial data generated by light-field imaging necessitates compression, careful consideration must be given to achieving an optimal balance among compression efficiency, reconstructed image quality, and the preservation of specific features required for accurate biomedical analysis. One critical factor is sample sparsity. As highlighted by Wang et al., compressed light-field sampling can result in reduced spatial resolution compared to complete sampling when the sparsity condition is not satisfied—for instance, when imaging complex biological structures or densely populated cell ensembles. Another important aspect is the preservation of disparity information. Recent research has shown that compressing the light field can adversely affect disparity estimation, thereby compromising depth accuracy [107]. Specifically, the authors demonstrated that applying state-of-the-art compression algorithms to light-field datasets, despite producing visually imperceptible differences between compressed and uncompressed perspective images, can negatively impact depth estimation and subsequent computer-vision-based tasks. Therefore, careful retention of disparity information during compression is essential. This principle is exemplified by the hardware implementation of SLIM, where the compression axis is intentionally aligned perpendicular to the disparity axis across different views, thereby maximizing the preservation of disparity information during light-field compression.

Fully harnessing the speed advantage of light-field imaging requires a synergistic approach that combines advancements in camera sensor technology with optimized strategies for light-field compression. Comparatively, compression strategies may yield more immediate improvements, given that the development cycle for camera hardware is typically lengthy. A promising future direction involves tailoring light-field compression specifically to target biological imaging applications through end-to-end optimization [109]. This approach requires retaining sufficient flexibility in the imaging hardware to allow optimization tailored directly to end applications. Given that light-field imaging fundamentally represents an optical tomography technique that samples object information in the frequency domain, the ability to program both the view angles and the compression ratios of corresponding perspective images emerges as a particularly promising strategy for achieving enhanced imaging performance.

*4.2 Field of view, depth of field, and resolution tradeoffs*

In light-field imaging, the field of view (FOV) and depth of field (DOF) are typically optimized for the target application. In microscopy, for instance, the FOV is determined by the objective's field number and magnification, while the DOF is set by the sub-aperture size associated with each view. Because the finite camera pixels are divided between spatial and angular samples by the microlens array [8–10], increasing angular sampling for improved depth reconstruction inherently reduces spatial sampling within a given FOV—unless offset by using a larger-format camera sensor or compensated computationally through advanced reconstruction algorithms.

The snapshot 3D imaging capability of light-field imaging comes at an inherent cost. By dividing the entrance pupil of the imaging optics into multiple sub-apertures, light-field imaging sacrifices diffraction-limited lateral resolution to achieve resolution along the depth axis. Given a fixed-size entrance pupil, the total number of spatially resolvable spots—also referred to as the system's space-bandwidth product (SBP)—remains constant. Light-field imaging does not increase the SBP of an imaging system; instead, it redistributes the available SBP from a purely lateral plane into three-dimensional space—the product of spatially resolvable voxels along the $x$, $y$, and $z$ axes in light-field imaging remains equal to the original product of spatially resolvable pixels along the $x$ and $y$ axes in conventional 2D imaging.

Various strategies have been proposed to compensate for the lateral resolution loss inherent in light-field imaging. A commonly adopted approach involves using fewer lenslets to sample the entrance pupil of the imaging optics, ensuring each sub-aperture remains large enough to yield high-resolution images. This method is widely utilized in light-field microscopy, particularly for subcellular imaging. However, since the depth of field in light-field imaging is determined by the numerical aperture of each individual lenslet, reducing the number of lenslets to enhance lateral resolution inevitably decreases the overall imaging depth range and effective depth sampling. To alleviate this trade-off, lenslet arrays with multiple focal lengths, each optimized for different depth ranges, have been employed. Yet, because the number of perspective images captured at each depth range is correspondingly reduced, this approach leads to sparser sampling in the frequency domain, limiting its applicability primarily to samples with even greater sparsity.

Alternatively, another notable approach is to reintroduce scanning, leading to the development of scanning light-field microscopy [87]. In this method, the intermediate image is scanned across the microlens array, thereby restoring diffraction-limited lateral resolution. However, the introduction of scanning inherently reduces imaging speed, diminishing one of the primary advantages of light-field imaging over other high-speed imaging techniques, such as light-sheet microscopy, especially when capturing rapid biological dynamics.

To address this limitation, the same group later developed virtual-scanning light-field microscopy (VsLFM), which replaces physical scanning with a physics-guided deep learning method [46]. By leveraging phase correlations among angular views, their virtual scanning network (Vs-Net) achieved lateral and axial resolutions of approximately 230 nm and 420 nm, respectively, comparable to the diffraction limit. By eliminating the need for physical scanning, VsLFM enables 3D imaging at the native frame rate of the camera, achieving acquisition speeds of up to 500 volumes per second. The authors demonstrated the effectiveness of VsLFM through high-speed volumetric imaging of various biological processes, including cardiac dynamics in embryonic zebrafish, voltage activity in Drosophila brains, and neutrophil migration in the mouse liver. Nonetheless, the downside of employing a deep neural network to replace physical scanning is its reliance on training data, which limits its applicability to samples similar to those encountered during the training phase. This requirement inherently restricts the model's generalizability to diverse biological samples.

Ultimately, resolving the resolution dilemma inherent in light-field imaging is challenging, as current strategies typically necessitate sacrificing other key benefits, such as high-speed acquisition or extended imaging depth. Rather than continually seeking to overcome these intrinsic limitations, it may be more productive to acknowledge and adapt to them. By accepting the resolution constraints of light-field imaging, researchers can more effectively leverage its unique advantages, such as rapid volumetric imaging and snapshot acquisition, clearly positioning its value within the existing biomedical imaging toolbox. Recognizing that no single imaging technology can meet every requirement, embracing the strengths and limitations of each modality allows for a more strategic and complementary approach to addressing diverse biomedical imaging needs.

*4.3 Meta-optics-powered light-field imaging*

Meta-optics refers to the control of light using artificially engineered materials, typically implemented as planar metasurfaces composed of subwavelength-scale elements [110]. Unlike conventional optical components (e.g., lenses, mirrors) that rely on surface curvature to geometrically bend light, meta-optics directly impart position-dependent phase shifts to the incident wavefront. This enables precise control over various optical properties, such as amplitude, phase, spectrum, and polarization. Such capabilities allow accurate capture and reproduction of multidimensional optical information beyond conventional spatial and angular dimensions (**Fig. 10**). Consequently, meta-light-field technology has the potential to address and improve upon the performance trade-offs typically encountered in traditional lenslet-based imaging systems.

Unlike traditional light-field imaging systems that utilize lens arrays to demultiplex incoming rays, meta-optics-powered light-field (meta-light-field) imaging employs metasurfaces composed of subwavelength nanostructures to precisely control various properties of incident light, such as amplitude, phase, spectrum, and polarization. This capability allows accurate capture and reproduction of multidimensional optical information beyond conventional spatial and angular dimensions (**Fig. 10**). Consequently, meta-light-field technology has the potential to address and improve upon the performance trade-offs typically encountered in traditional lenslet-based imaging systems.

For example, Park et al. introduced a virtual-moving metalens array (VMMA) that enhances the resolution of light-field imaging [111]. This method is conceptually similar to scanning light-field microscopy but achieves improved sampling without physically translating the image across the microlens array. Fabricated as a polarization-sensitive phase device, the VMMA produces different lateral focal positions depending on the polarization of the incident light. By switching the polarization of the illumination beam, the periodic sampling positions of the VMMA shift accordingly, effectively doubling the spatial sampling density of the light field. However, with scanning light-field microscopy, the need to alternate the polarization state of the illumination slows down the overall imaging speed. Moreover, this polarization-dependent approach is limited to transmission imaging, and it requires that the sample does not alter the polarization state of the incident light, restricting its applicability in complex or birefringent biological samples.

In another example, Holsteen et al. developed a light-field metasurface by interleaving three metalenses within a single 200 μm-diameter shared aperture, using the Pancharatnam–Berry phase to encode different phase profiles [102]. The optical axes of the metalenses are linearly spaced with a separation of 66 μm, allowing each to image the same 3D scene from different view angles. By multiplexing the phase profiles onto a shared aperture, the effective aperture for each view equals the full clear aperture of the array, thereby enhancing spatial resolution without compromising angular resolution—or, equivalently, effective depth sampling. However, this phase multiplexing approach can introduce optical crosstalk between views, posing challenges for scaling the design to configurations with a larger number of metalenses.

Additionally, meta-optics has been explored as a means to alleviate the trade-off between lateral resolution and depth of field in light-field imaging. In conventional lenslet-based systems, increasing the aperture of each lenslet improves lateral resolution but reduces the depth of the field. To overcome this limitation, one meta-light-field approach employed polarization multiplexing [112] to achieve dual focal points—similar in concept to lenslet arrays with dual focal lengths, as discussed in a previous section—but without reducing sampling density in the frequency domain, since both focal lengths share the same sub-aperture. However, this metalens system suffers from spatially non-uniform optical aberrations, including chromatic and geometric aberrations introduced by the metasurface, which degrade lateral resolution beyond the diffraction limit—a common issue in metalenses designed for an extended depth of field. To address this problem, the authors developed a multi-scale convolutional neural network-based reconstruction algorithm to correct for these aberrations. This integration of meta-optics with computational imaging enables full-color light-field imaging with a significantly extended depth of field, ranging from 3 cm to infinity in a photographic setup.

Moreover, because metasurfaces are 2D thin structures, they can be directly integrated into camera sensors, facilitating the creation of highly compact and integrated imaging chips. For instance, Lin et al. developed the first full-color, achromatic light-field camera utilizing a metalens array made from gallium nitride nanoantennas [113]. This metalens array comprises 60 × 60 individual metalenses, each with a diameter of 21.65 μm. Each metalens within the array is free of chromatic aberration and achieves diffraction-limited resolution with a low F-number. The utility of this sensor has been successfully demonstrated through full-color 3D imaging experiments.

Despite their great promise, the optical performance of current metalenses still falls short of that of traditional compound lenses, particularly in large field-of-view imaging. While the ultimate capabilities of meta-light-field systems hinge on continued advances in metasurface design and fabrication, integrating meta-light-field imaging with computational techniques offers a promising intermediate solution. This hybrid approach can help compensate for optical aberrations and enhance overall imaging performance, accelerating the practical adoption of metasurface-enabled light-field technologies.

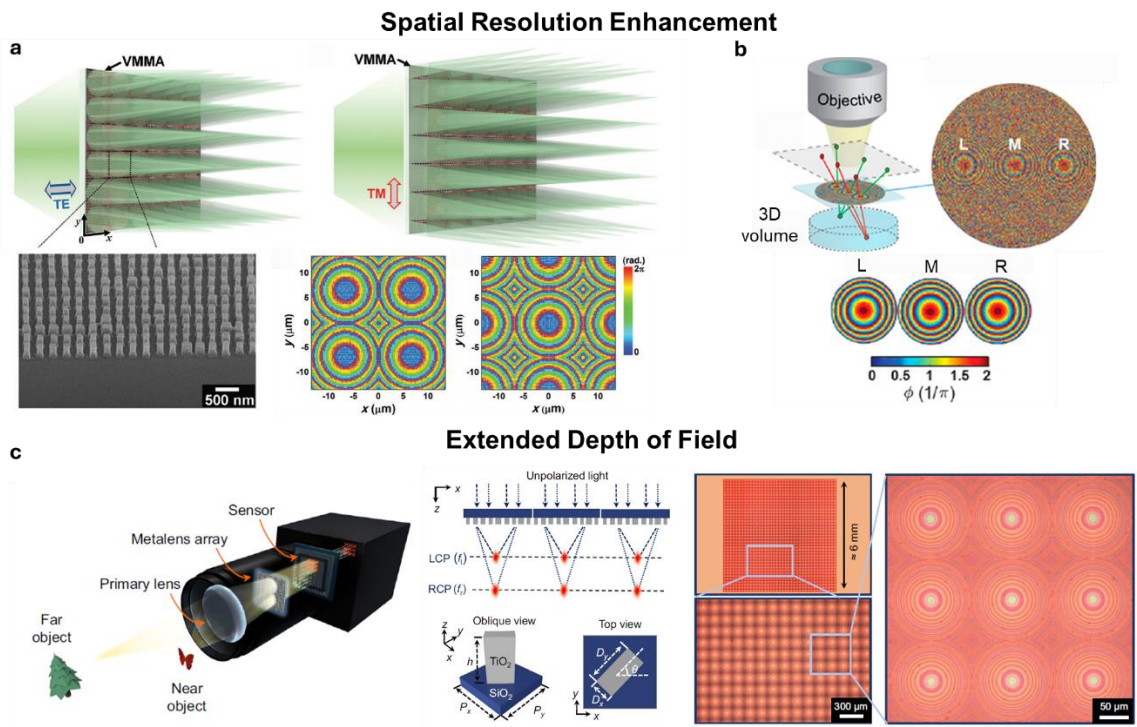

**Figure 10.** Meta-optics enhanced light-field imaging system. (a) Resolution enhancement with a virtual moving microlens array. (b) Resolution enhancement with Synthetic aperture. (c) Extended depth of field with a metalens array. Figures are reprinted with permission from Ref. [111], Wiley-VCH. Ref. [112], American Chemical Society, and Ref. [113], Springer Nature.

### 4.4 AI-powered light-field imaging

Light-field imaging is fundamentally a limited-view optical tomography technique, and as such, it suffers from the well-known "missing cone" problem and sparse sampling in the frequency domain. Traditional light-field reconstruction often relies on iterative algorithms, such as Richardson–Lucy deconvolution, to generate 3D images. However, these methods are computationally intensive and prone to artifacts in reconstruction. Deep learning has recently emerged as a transformative solution to these challenges. Recent studies have demonstrated that deep learning–based reconstruction methods effectively address persistent issues in light-field imaging, including low and non-uniform resolution and high computational cost (**Fig. 11**). Both fully end-to-end neural network models and hybrid approaches—where physical inversion techniques are combined with deep learning enhancements—now enable fast, robust, and artifact-free reconstructions. Comprehensive reviews, including that by Lin et al. [114], have detailed the rapid evolution and impact of deep learning in advancing light-field reconstruction methods.

Deep-learning-based light-field reconstruction methods can generally be categorized into two main approaches: fully deep learning-based methods and hybrid methods combining deep learning with numerical inversion. Fully deep learning-based methods involve training neural networks to directly convert raw light-field data into high-resolution 3D volumes [64,115–117]. These models learn the mapping from low-resolution, artifact-prone inputs to refined 3D reconstructions, dramatically reducing computation time [117]. For example, Vizcaino et al. [115] introduced LFM-Net, an innovative approach that reconstructs confocal microscopy stacks from a single light-field image. This method offers significant improvements in both reconstruction accuracy and speed over earlier techniques. However, a major limitation of fully deep learning-based approaches is their generalization: models are typically trained on specific types of objects or scenes and may perform poorly when applied to unfamiliar data.

In contrast, hybrid methods integrate deep learning at one or more stages of the reconstruction pipeline, leveraging both data-driven learning and physics-based modeling [46,48,118–124]. One strategy applies deep learning to enhance the resolution of raw light-field images prior to numerical inversion. Another approach performs an initial iterative reconstruction to produce a low-quality 3D volume, which is subsequently refined using a neural network. For example, Qiao et al. [118] introduced architectures such as the Deep Fourier Channel Attention Network (DFCAN) and its generative-adversarial-network (GAN)-augmented variant (DFGAN), which employ frequency-domain attention mechanisms to recover fine details under challenging imaging conditions. By combining the robustness of numerical inversion with the adaptability of deep learning, hybrid approaches offer improved generalization and reconstruction quality. Additional strategies include learned iterative shrinkage and thresholding algorithms (LISTA) [119], which simplify system models and reduce artifacts, as well as unsupervised approaches using GANs to enable reconstruction without the need for labeled training data.

Unlike conventional deep learning approaches used to enhance 2D biomedical images, the design of neural network architectures for light-field reconstruction must account for the unique structure of light-field data, which inherently encodes both spatial and angular information. One notable example is the View-Channel-Depth (VCD) model [117,121], which transforms raw 2D light-field images by reorganizing microlens "views" into multi-channel tensors—effectively mapping angular information into separate channels. This view-channel-depth transformation enables the network to learn correlations between spatial and angular cues, facilitating accurate depth inference. For instance, Yi et al. [64] introduced a VCD-based framework (F-VCD) that reconstructs high-resolution 3D volumes from 2D light-field snapshots. These models capitalize on the intrinsic relationship between angular diversity and scene depth, achieving notable improvements in both reconstruction resolution and computational speed.

For time-lapse imaging over extended periods, it is essential that reconstructions remain accurate and reliable across varying experimental conditions. To address this, continuous validation frameworks have been developed, utilizing high-resolution reference images—often acquired through complementary modalities, such as light-sheet microscopy—to assess and refine the neural network's output in real-time. Wagner et al. [116] demonstrated such an approach with HyLFM-Net, a system that integrates simultaneous high-resolution light-sheet data into the light-field reconstruction pipeline. This dual-modality setup enables on-the-fly retraining or fine-tuning of the neural network, ensuring consistent reconstruction quality even as imaging conditions evolve over the course of long-term experiments.

Deep learning–enhanced light-field imaging has rapidly evolved into a practical and powerful tool for addressing complex biological and medical questions. One of the most exciting applications lies in dynamic imaging of living cells. For example, approaches based on VCD networks have enabled video-rate 3D imaging of intracellular processes. In [64], the authors demonstrated that this method could capture fast intracellular events such as mitochondrial fission and fusion in real time.

Despite these advancements, significant challenges remain. Most current deep-learning-based light-field reconstruction methods rely on supervised learning, which requires substantial training datasets and is prone to overfitting. The generalizability of these models across different imaging setups and biological conditions remains an open question. Additionally, many deep learning models function as black boxes, with limited interpretability of their internal decision-making processes. Their performance heavily depends on the quality and diversity of the training data, and there are few guarantees as to when a model will accurately enhance the image versus when it might introduce hallucinated or misleading information. Ensuring accountability, interpretability, and reliability in deep-learning models for reconstructing light-field images remains a critical challenge in the field.

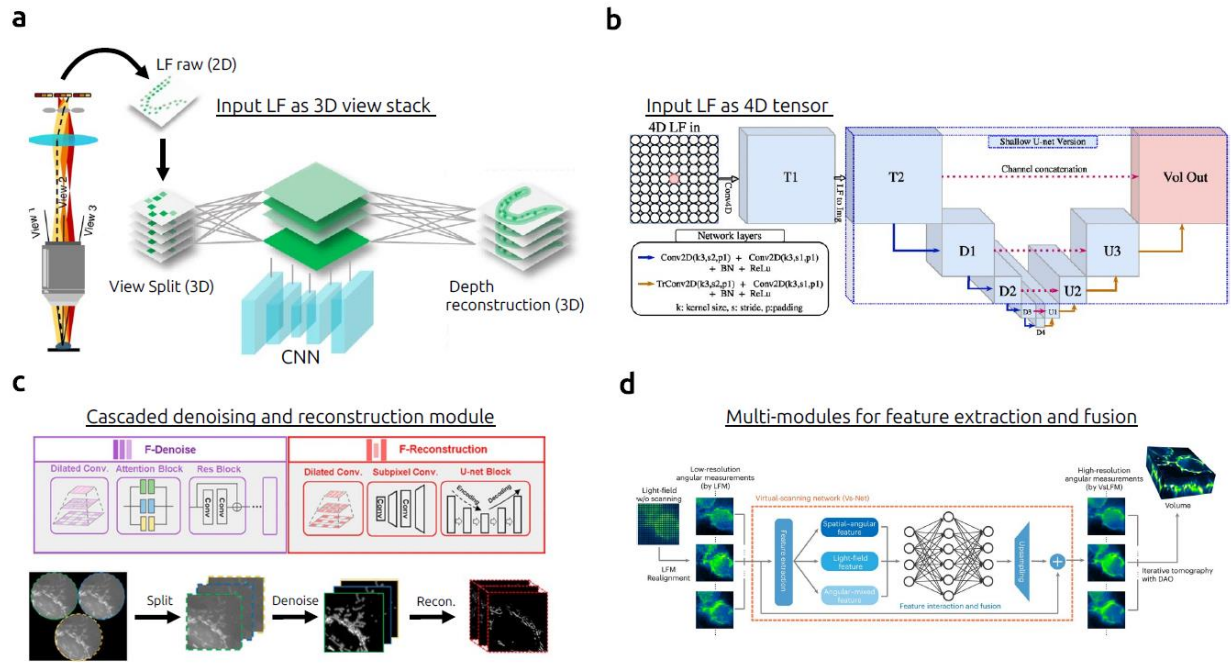

**Figure 11.** Deep learning for image reconstruction in light-field microscopy. (a) VCD-Net interfaces with a light-field view stack and outputs a 3D image. (b) LFMNet begins with a 4D convolutional layer to process the original light field in 4D. (c) F-VCD extends VCD-Net to Fourier LFM and improves it with a two-stage network for both denoising and reconstruction. (d) VsLFM is trained to map data between LFM and scanning LFM, utilizing multiple modules for feature extraction and fusion. Figures are reprinted with permission from Refs. [46,64,118], Springer Nature, and Ref. [115], IEEE.

## 5. Declarations

**Ethics Approval**

Not applicable

**Acknowledgement**

This work was supported by the following grants: National Institutes of Health (NIH) (R01HL165318, RF1NS128488, R35GM128761, R01AI102584, R01HL129727, R01HL159970, 5T32HL144449).

**Abbreviations**

Three dimensional(3D); two-dimensional (2D); signal-to-noise ratio (SNR); seven-dimensional (7D); four-dimensional (4D); numerical-aperture (NA); field of view(FOV); depth of field(DOF); light-field microscopy (LFM); fourier light-field microscopy (FLFM); extended light-field microscopy (XLFM); space-bandwidth product (SBP); mesoscale selective plane illumination microscopy (MesoSPIM); mesoscale light-field (MesoLF); computational miniature mesoscope (CM2); real-time, ultra-large-scale, high-resolution 3D mesoscope (RUSH3D); multiscale background rejection (MBR); gradient refractive index(GRIN); liquid-crystal (LC); multifunctional LC lens (MFLC-lens); green fluorescent protein (GFP); squeezed light-field microscopy (SLIM); virtual-scanning light-field microscopy (VsLFM); virtual scanning network (Vs-Net); virtual-moving metalens array (VMMA); deep fourier channel attention network (DFCAN); generative-adversarial-network (GAN)-augmented variant (DFGAN); learned iterative shrinkage and thresholding algorithms (LISTA); View-Channel-Depth (VCD); hybrid LFM (HyLFM);